\documentclass[twocolumn,twoside,slac_two]{revtex4}
\usepackage{graphicx}
\usepackage{fancyhdr}
\usepackage{longtable}
\newcommand{\bei}{\begin{itemize}}
\newcommand{\eei}{\end{itemize}}
\newcommand{\beq}{\begin{equation}}
\newcommand{\eeq}{\end{equation}}
\newcommand{\beqn}{\begin{eqnarray}}
\newcommand{\eeqn}{\end{eqnarray}}
\newcommand{\beqns}{\begin{eqnarray*}}
\newcommand{\eeqns}{\end{eqnarray*}}

\newcommand{\mc}{\multicolumn}

\newcommand\rs{\raisebox{1.3ex}[-1.5ex]}

\newcommand{\MSbar}{\overline{\rm MS}}

\newcommand\ea{{\em et al.}}

\newcommand\Amptpbar{\kern 0.18em\overline{\kern -0.18em {\cal A}}_{3\pi}}

\newcommand\Amptpbarkappa{\kern 0.18em\overline{\kern -0.18em A}^{\kappa}{}}
\newcommand\Amptpbarsigma{\kern 0.18em\overline{\kern -0.18em A}^{\sigma}{}}

\newcommand\Kzpiznn{{\cal B}(K^0_{\rm L}\to\pi^0\nu\nub)}
\newcommand\Kppipnn{{\cal B}(K^+\to\pi^+\nu\nub)}
\newcommand\Bpmun{{\cal B}(B^+\to\mu^+\nu_\mu)}
\newcommand\Bptaun{{\cal B}(B^+\to\tau^+\nu_\mu)}
\newcommand\dmd{\Delta m_d}

\newcommand\dms{\Delta m_s}

\newcommand\epsk{\varepsilon_K}

\newcommand\Nbpm{{\kern 0.18em\overline{\kern -0.18em N}}^{+-}}
\newcommand\Nbmp{{\kern 0.18em\overline{\kern -0.18em N}}^{-+}}

\newcommand\stbwa{\sin(2\beta)_{[c\bar c]}}

\newcommand\BRpmb{{\cal \kern 0.18em\overline{\kern -0.18em  B}}{}_{\rho\pi}^{+-}}
\newcommand\BRmpb{{\cal \kern 0.18em\overline{\kern -0.18em  B}}{}_{\rho\pi}^{-+}}

\newcommand\BRipmb{{\cal \kern 0.18em\overline{\kern -0.18em  B}}{}_{\rho^+\pi^-}}
\newcommand\BRimpb{{\cal \kern 0.18em\overline{\kern -0.18em  B}}{}_{\rho^-\pi^+}}

\newcommand\Abar{\kern 0.18em\overline{\kern -0.18em A}{}}

\newcommand\mcRun{{\overline m}_c}

\newcommand\mtRun{{\overline m}_t}
\newcommand\rfit{{\em R}fit}
\newcommand\etaB{\eta_B}

 \input myBabarsym

\begin{document}

\title{Constraints on the CKM Matrix}


\author{J\'er\^ome Charles (for the CKMfitter group)}
\affiliation{Centre de Physique Th\'eorique, Luminy Case 907, 13288 Marseille Cedex 9, France}

\begin{abstract}
We update the analyses of the Cabibbo-Kobayashi-Maskawa matrix, both
within the Standard Model and for arbitrary New Physics contributions
to the mixing amplitudes, using new inputs from the Winter 2006
conferences. 
\end{abstract}

\maketitle

\thispagestyle{fancy}


\section{Introduction}

The most important observables that were part of the planned $B$-factory
program have now been
measured. With the addition of the new $\dms$ constraint from the Tevatron experiments, the three main
flavor-changing neutral current transitions ($s\to d$, $b\to d$ and $b\to s$) are tested to different precision levels and
compared to Standard Model predictions. In these proceedings we update the analyses of the
Cabibbo-Kobayashi-Maskawa (CKM) matrix of Ref.~\cite{bible} (where all 
notations and
technical details can be found) with the recently measured relevant observables.

\section{Inputs to the global CKM fit}
The inputs are listed in Tables~\ref{tab:ckmInputs} and~\ref{input2}. In the following we discuss to some detail the
status of the angles $\alpha$ and $\gamma$, and of the $B_s\bar{B_s}$ oscillation frequency $\dms$.

\subsection{The angle \boldmath$\alpha$}
\begin{figure}[h]
\includegraphics[width=0.475\textwidth]{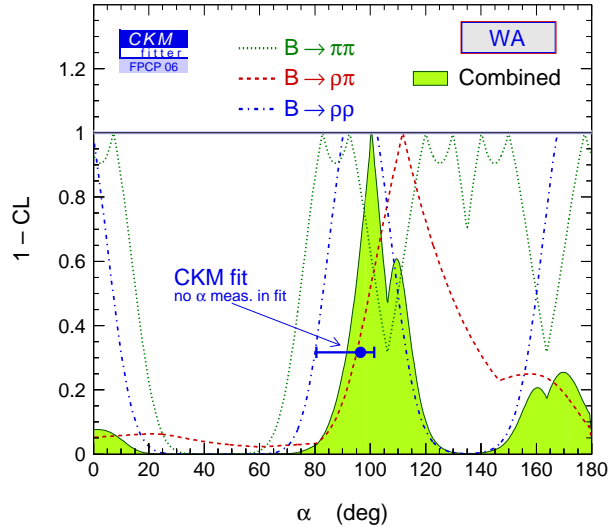}
\caption{Confidence level for the angle $\alpha$ from $B\to\pi\pi$,
$B\to\rho\pi$,
$B\to\rho\rho$ and combination (world average data).}\label{alpha1D}
\end{figure}

The current direct constraint on $\alpha$ comes from mixing-induced CP-violating measurements,
through the combination of the two-body isospin
analyses of $B\to\pi\pi$ and $B\to\rho\rho$, and the Dalitz plot analysis of
$B\to\rho\pi$ (see, e.g., Ref.~\cite{babarbook} and references therein). Among the three channels
$B\to\pi\pi$ plays only a minor r\^ole because of the pattern of discrete ambiguities. In
$B\to\rho\rho$ a missing observable, namely BR$(B\to\rho^0\rho^0)$, prevents to perform a full
Gronau-London~\cite{GL} analysis.  The current upper bound on this mode actually implies bounds \`a
la Grossman-Quinn~\cite{GQ} on
the difference $|\alpha-\alpha_\mathrm{eff}|$. Before the update on the $B\to\rho^+\rho^0$ branching fraction that was
presented by the \babar\ collaboration at the winter conferences~\cite{alpha}, the world average
data were in slight disagreement with the existence of an isospin
triangle~\cite{bible}. This somewhat ``lucky''
fluctuation was reflected in the fit result by a sharp peak, that has now evolved to a plateau and a
bit larger error on $\alpha$, in agreement with what is expected from theoretical bounds. We find
at 68\% CL (Fig.~\ref{alpha1D}) 
\beq
\alpha = (100^{+15}_{-9})^\circ\,.
\eeq

\subsection{The angle \boldmath$\gamma$}

\begin{figure}[h]
\includegraphics[width=0.475\textwidth]{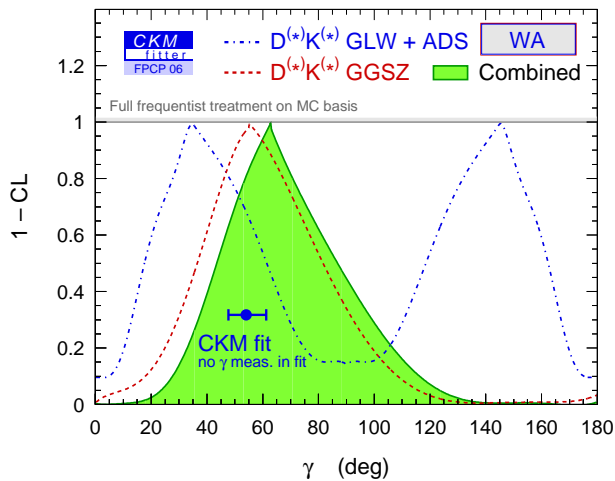}
\caption{Confidence level for the angle $\gamma$ from the GLW, ADS and
GGSZ methods~\cite{gamma}
(world average data).}
\label{gamma1D}
\end{figure}

The extraction of $\gamma$ stems from direct CP-violation measurements in $B\to DK$ modes.
Although it is theoretically simpler than that of $\alpha$, because the formulae are
more compact and do not involve the use of flavor symmetry~\cite{gamma,lees}, the statistical
interpretation of present data requires advanced techniques. Of crucial importance for the performance of the analysis
is the size of
the $r_B$ parameter (where there is one for each decay channel), the ratio of $b\to u\bar cs$ to $b\to c\bar us$
amplitudes:
the larger $r_B$, the smaller the error on $\gamma$. With current statistics however, $r_B$ remains
not too far from zero, which in turn implies that the minimum $\chi^2$ result for $r_B$ (resp. the
error on $\gamma$) is biased towards larger values (resp. smaller values). In order to evaluate this
bias and to correct for this unwanted effect, one must perform a full frequentist analysis by
means of toy Monte-Carlo studies as a function of the true parameter values. Both
\babar~\cite{babargamma}  and Belle~\cite{bellegamma}
use a Neyman-type construction of the confidence level in the full parameter space, with different
choices of the ordering function~\cite{yabsley} (Belle's choice is
equivalent to the substraction of the global
minimum from the $\chi^2$  function). Then the next question is how to get rid of the nuisance
parameters ($r_B$ and the associated strong phase) in order to determine the confidence level for the
desired parameter $\gamma$. Both \babar\ and Belle make a gaussian-like assumption by using the
known correspondence between confidence levels of different dimensionality, through the
specification of the number of degrees of freedom~\cite{PDG32}. While this is presumably a very good
approximation for the case of $\gamma$, this method is not completely general as in some situations
the number of degrees of freedom of the likelihood function can be
ill-defined. For the sake of generality the
CKMfitter group has decided to use a new method that avoids this
assumption. Its application to the
present case leads to a slightly larger error on $\gamma$: for the \babar\
data on $B\to
DK$ in the Dalitz plot analysis (statistical errors only) CKMfitter finds a 68\%
CL interval of $[35^\circ,102^\circ]$ instead of 
$[39^\circ,101^\circ]$ as quoted by \babar. The bad news is that even when one
averages over all channels and all data one finds the rather loose
determination~\footnote{With the frequentist method that was used by
CKMfitter for the summer 2005 conferences (with the result $\gamma =
(63^{+15}_{-12})^\circ$
~\cite{web}), but with the present winter 2006 data, we find $\gamma =
(62^{+28}_{-19})^\circ$. This  method is being generalized.}
(Fig.\ref{gamma1D})
\beq
\gamma = (62^{+35}_{-25})^\circ\,.
\eeq
More detailed studies on the origin of this error (presumably a combination of statistical effects
and preferred $r_B$ values), as well as coverage tests (preliminary results show a reasonably good
behavior) will be published elsewhere.

\subsection{The oscillation frequency \boldmath$\dms$}

\begin{figure}[h]
\includegraphics[width=0.475\textwidth]{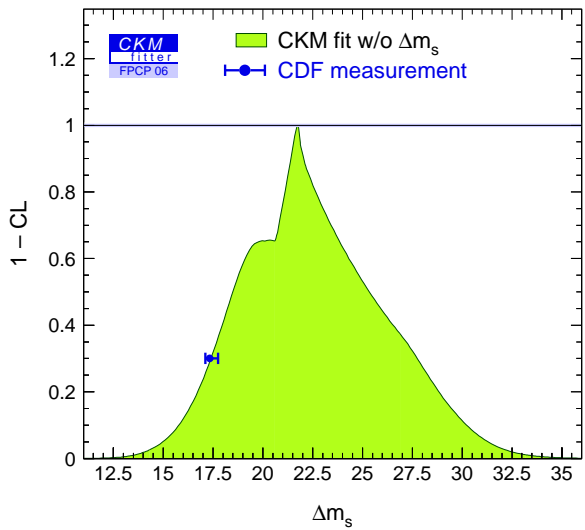}
\caption{Comparison of the direct measurement of $\dms$ by the CDF collaboration with the indirect prediction
from the standard global fit.}\label{dms}
\end{figure}

The most important news concerning flavor physics at the 2006 winter conferences is of course the
first two-sided bound on $\dms$ by D0~\cite{D0} followed by the $99.5\%$ SL measurement by CDF~\cite{CDF}. The
68\% CL range found by CDF, $\dms=17.33^{+0.42}_{-0.21}\pm 0.07\ \mathrm{ps}^{-1}$, is in excellent
agreement with the indirect prediction of the Standard Model global fit, $\dms=21.7^{+5.9}_{-4.2}\ \mathrm{ps}^{-1}$, as can be seen in
Fig.~\ref{dms}.
Taken as an input, the combination of $\dms$ and $\dmd$ has a strong impact on the
$(\rhobar,\etabar)$ plane (Fig.~\ref{rhoeta}), but the corresponding constraint is now completely
dominated by the theoretical uncertainty (from lattice simulations) on the ratio $\xi$ of the
relevant matrix elements.

\subsection{Theoretical uncertainties}

\begin{figure*}[Ht]
\includegraphics[width=0.475\textwidth]{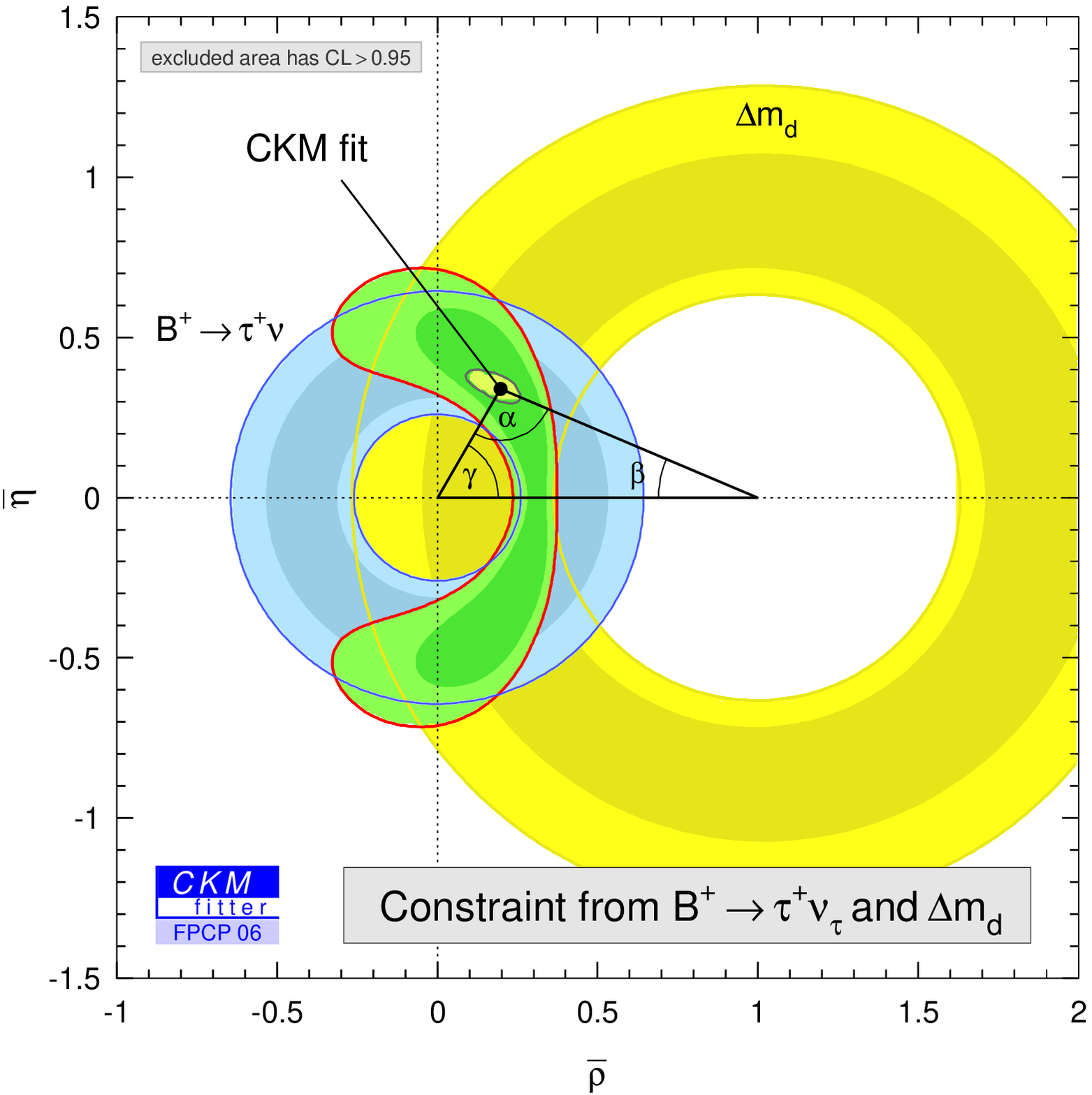}\hfill%
\includegraphics[width=0.475\textwidth]{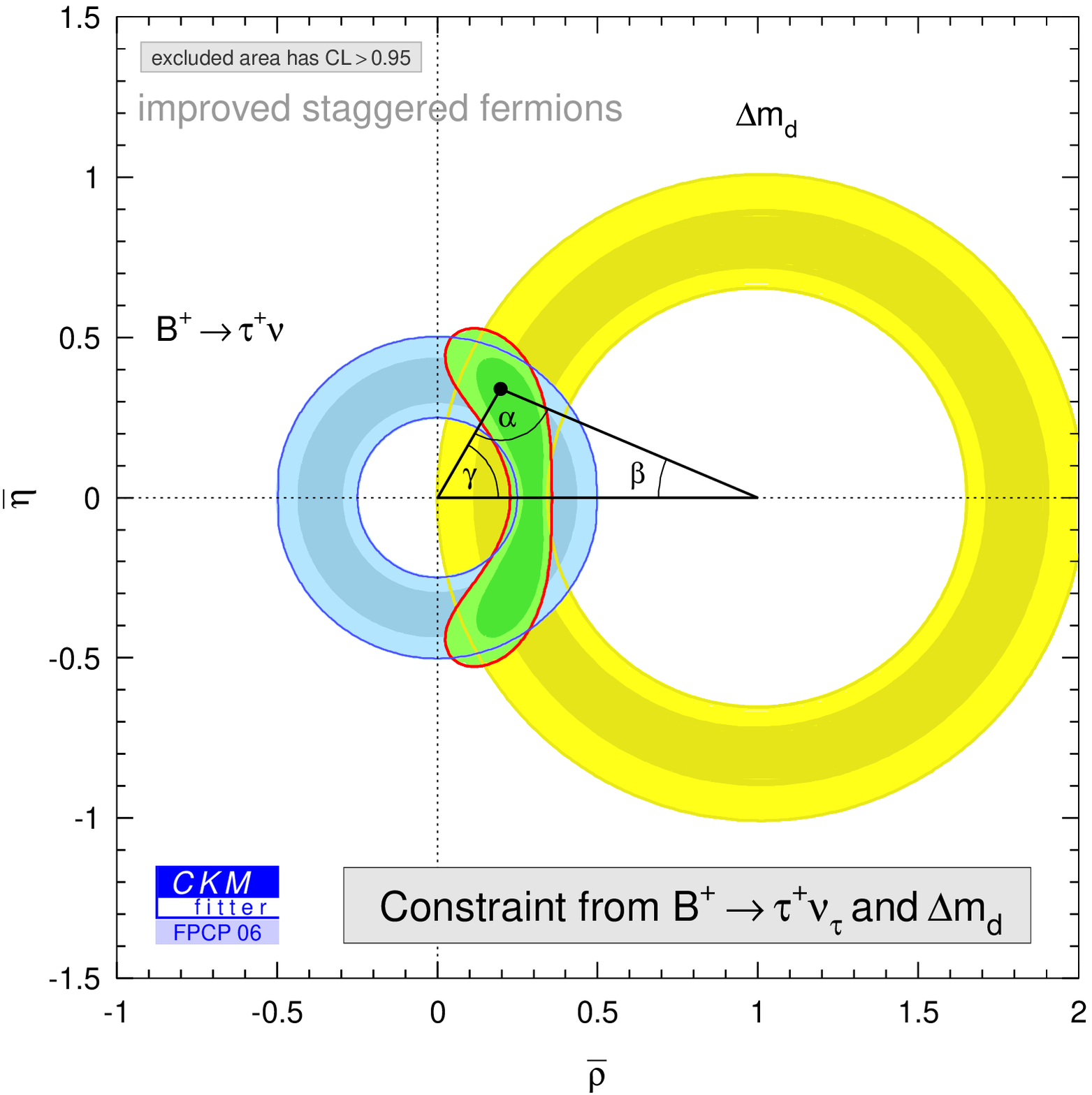}
\caption{95\% CL constraints in the $(\rhobar,\etabar)$ plane from
$B\to\tau\nu$, $\dmd$ and combination. In
the combination the decay constant $f_{B_d}$ cancels, so that the only
source of theoretical
uncertainty comes from the bag factor $B_{B_d}$. Left: our default input
 values (Tables~\ref{tab:ckmInputs} and~\ref{input2}); right: inputs
from improved staggered fermions
 formalism~\cite{oka}.}\label{theo}
\end{figure*}
All non-angle masurements are now dominated by theoretical uncertainties, coming from the evaluation
of non perturbative hadronic effects by lattice simulations, QCD sum rules or possibly other
methods. Nevertheless these errors have decreased and further significant progress is expected. 
Fig.~\ref{theo} shows the constraints on $(\rhobar,\etabar)$ coming from $\dmd$ and the recent
Belle's measurement of the leptonic decay~\cite{btaunuFPCP}
BR$(B\to\tau\nu)=(1.06^{+0.34}_{-0.28}\mathrm{(stat)
}^{+0.22}_{-0.25}\mathrm{(syst)}) \times 10^{-4}$, and of the combination
of the two. The left plot
corresponds to our default input values, while for the right plot the results
obtained with
the improved staggered fermion action have been used. The latter method aims at the efficient estimate of
unquenched effects coming from the light sea quark masses~\cite{mack}. While the outcome of this
approach is impressive and has passed several non trivial tests, the precise relation to full
continuum QCD is not completely understood and there is no consensus on the accuracy of this
formalism.

As a side remark to this comparison, we would like to stress that there are significant
correlations between the evaluation of the various theoretical parameters, that have a r\^ole when
the errors decrease. For example the two choices
$(f_{B_d},f_{B_s}/f_{B_d},B_{B_s},B_{B_d})$ and
$(f_{B_d},f_{B_s}/f_{B_d},B_{B_s},B_{B_s}/B_{B_d})$ for the decay constants and bag parameters are
mathematically equivalent, but the corresponding predictions for the relevant observables are  different
if one uses the inputs of Ref.~\cite{oka}. This is because we have neglected correlations between
the parameters. We then encourage our colleagues working on the lattice to publish the correlation
matrices found in the simulations, at least for the purely statistical part of the error.

\section{Fit results}

\begin{figure}[Ht]
\includegraphics[width=0.475\textwidth]{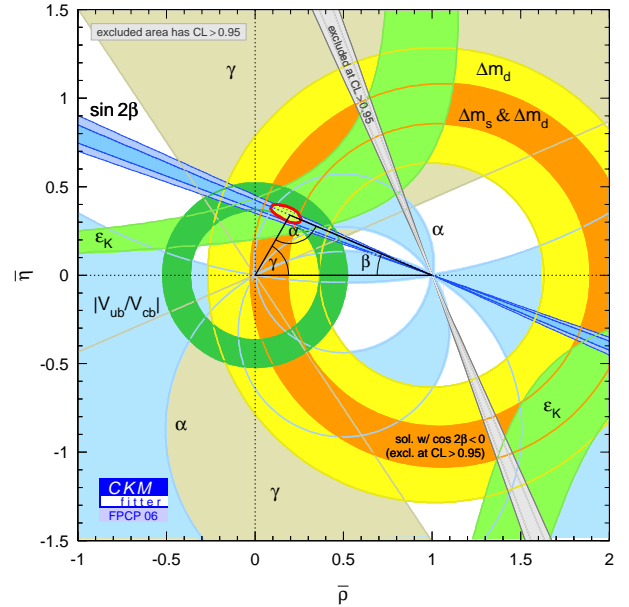}
\caption{Confidence level in the $(\rhobar,\etabar)$ plane for the global CKM fit. The shaded area
indicate the regions of at least $95\%$ CL.}\label{rhoeta}
\end{figure}

The global CKM reference fit is defined as the combination of constraints on the CKM matrix elements
on which we think we have a sufficiently good theoretical control. This correspond to the charged
current couplings $|V_{ud}|$, $|V_{us}|$, $|V_{cb}|$ as well as the following quantities that are
specifically sensitive to $(\rhobar,\etabar)$: $|V_{ub}|$, $|\epsilon_K|$, $\dmd$, $\dms$, $\alpha$,
$\beta$ and $\gamma$. The individual constraints as well as the combination are shown in
Fig.~\ref{rhoeta},
all of them are in good or even excellent agreement with each other. Tables~\ref{output}
and~\ref{output2} show
the $1\sigma$, $2\sigma$ and $3\sigma$ allowed ranges for several particularly interesting
parameters and observables.
\begin{figure*}[Htb]
\includegraphics[width=0.475\textwidth]{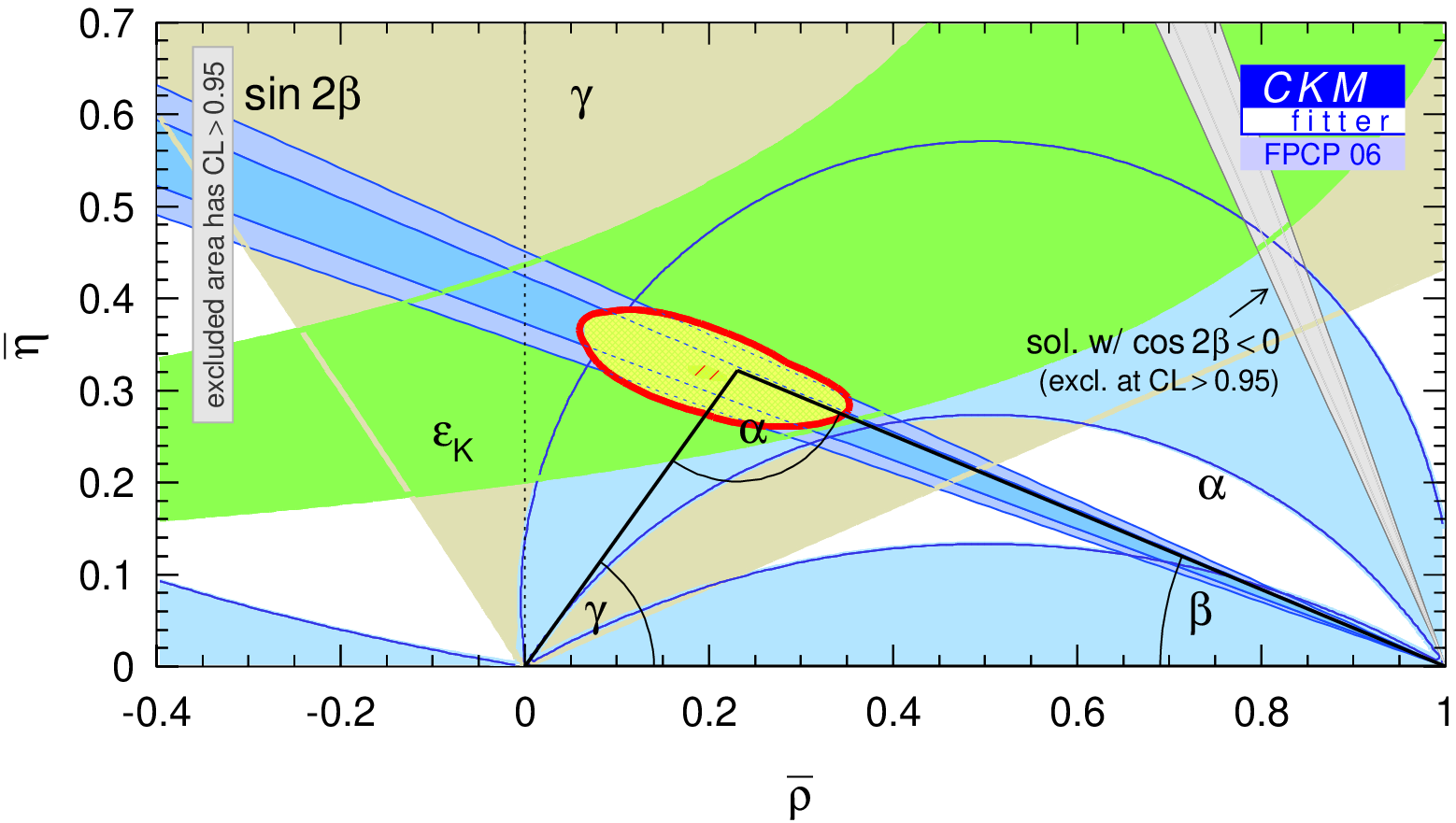}\hfill%
\includegraphics[width=0.475\textwidth]{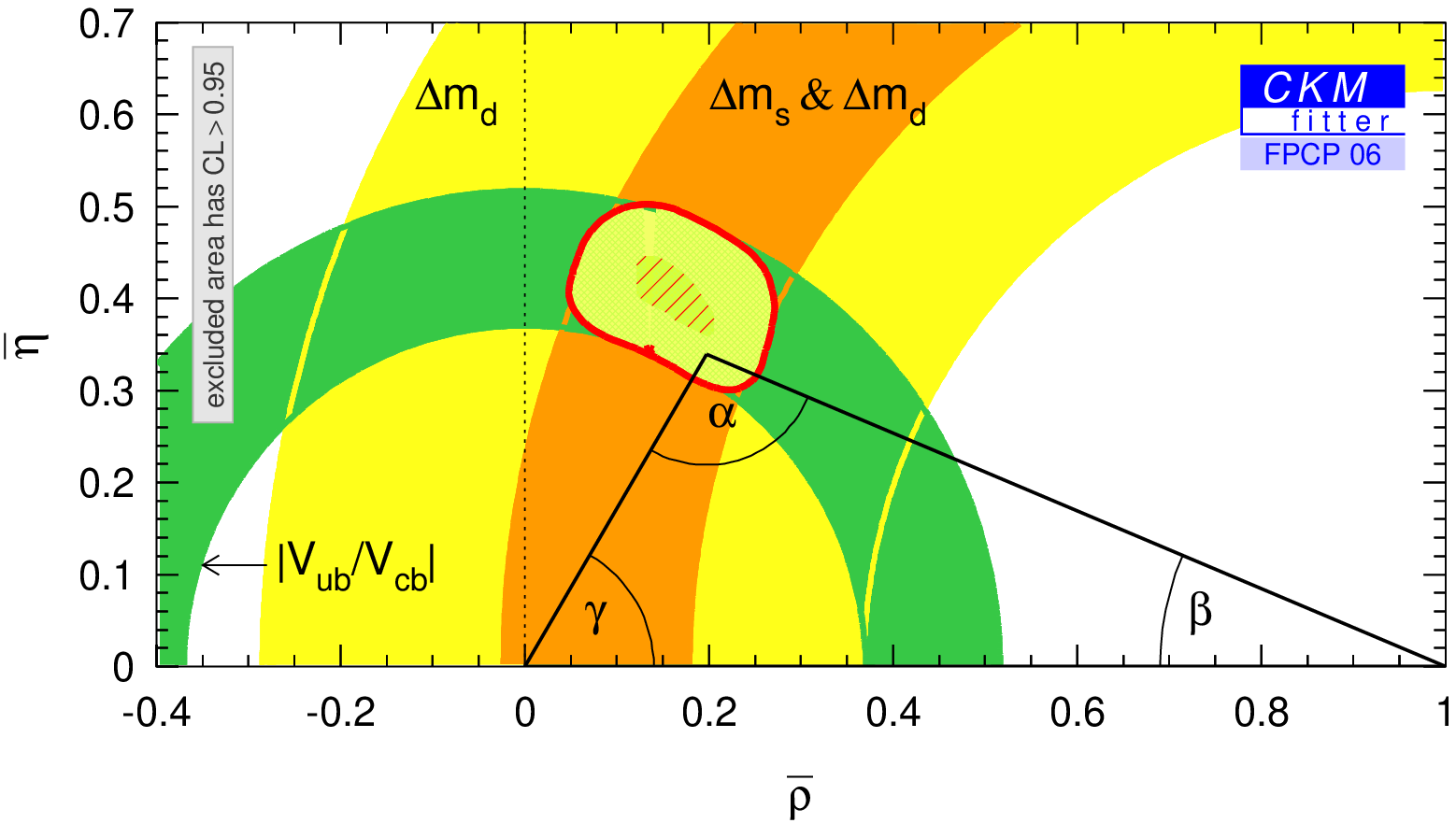}
\includegraphics[width=0.475\textwidth]{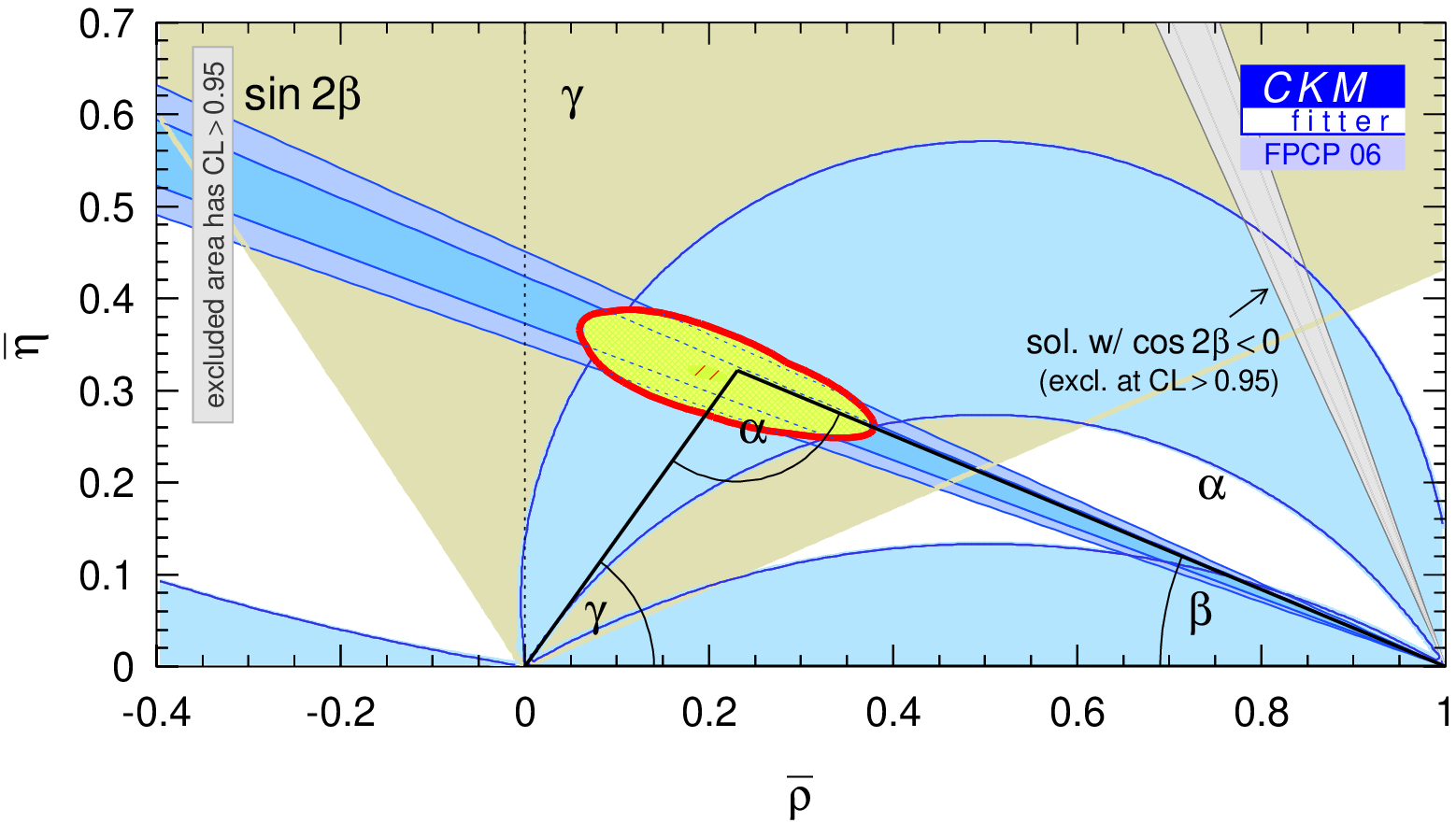}\hfill%
\includegraphics[width=0.475\textwidth]{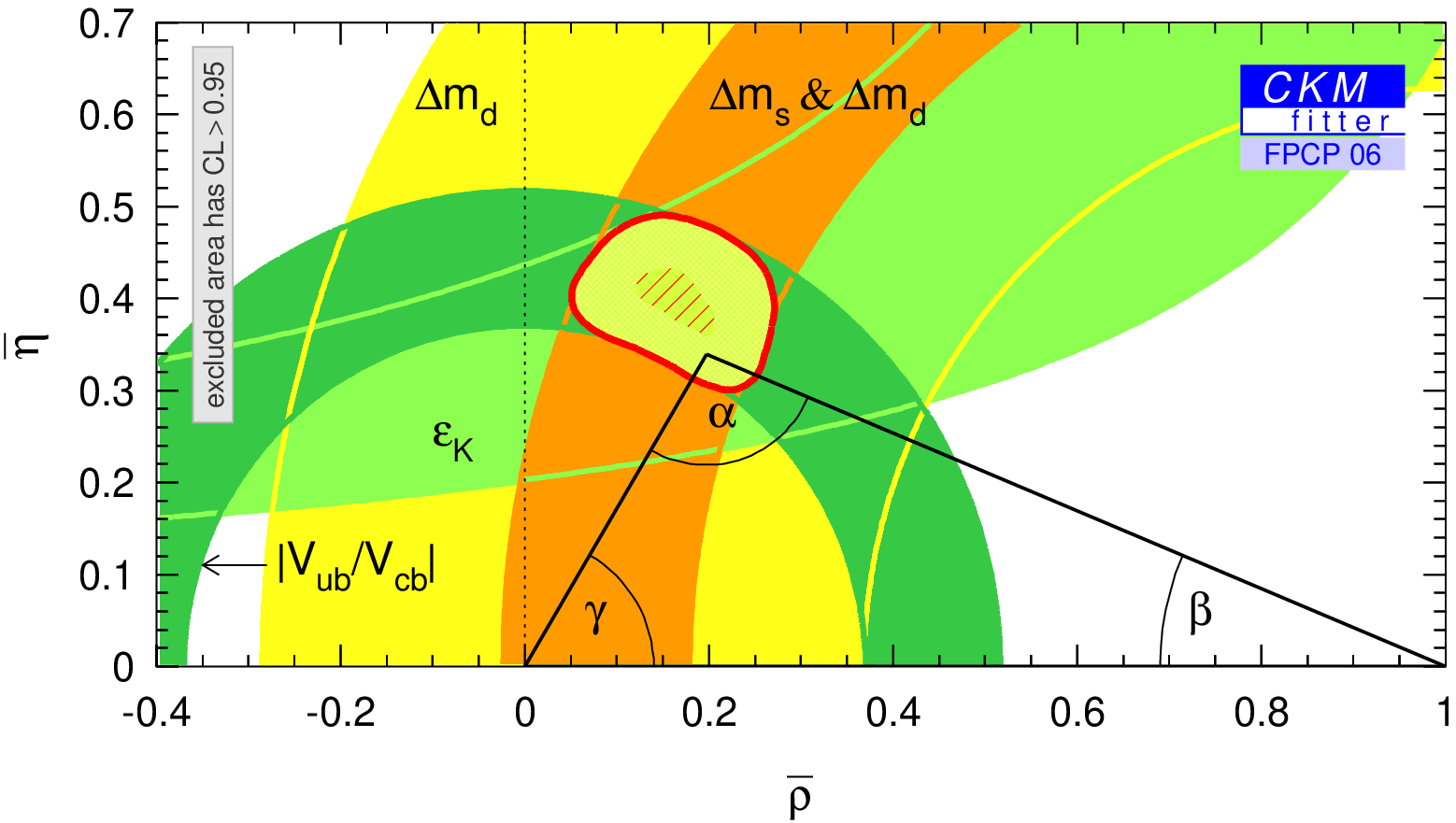}
\includegraphics[width=0.475\textwidth]{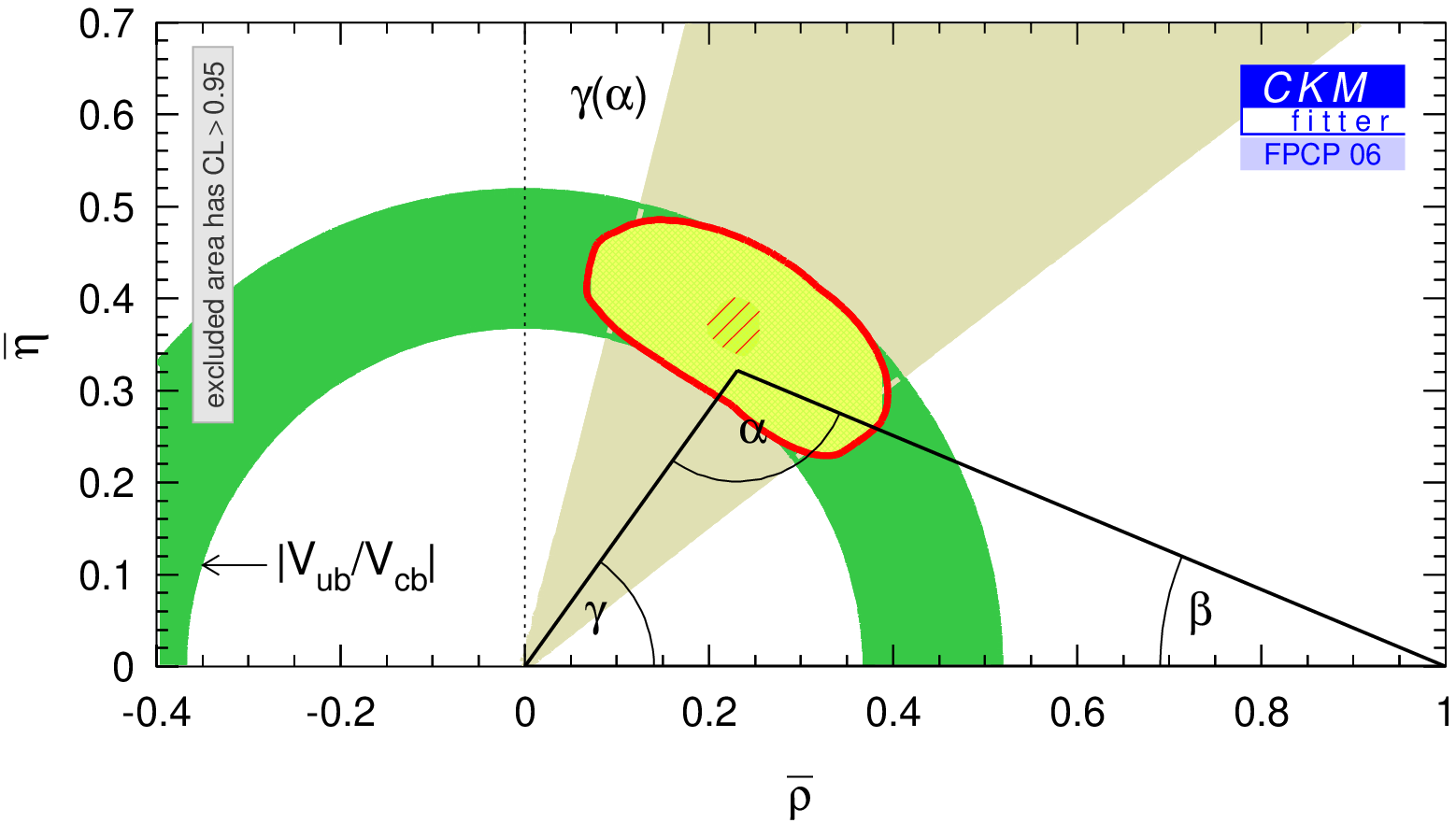}\hfill%
\includegraphics[width=0.475\textwidth]{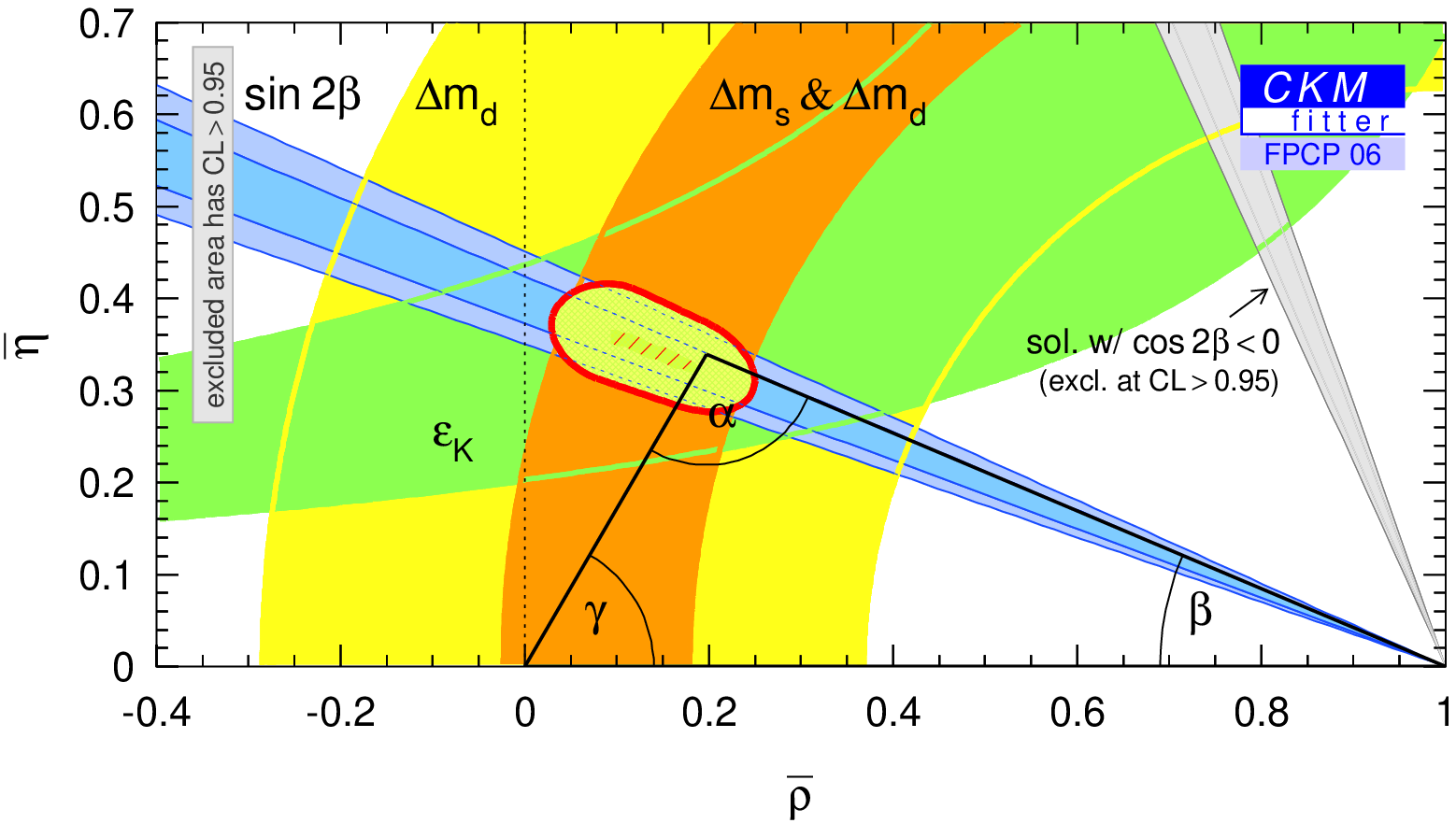}
\caption{Confidence level in the $(\rhobar,\etabar)$ plane for the
global CKM fit. From top to
bottom and left to right:
CP-violating observables vs. CP-conserving ones,
theory-free constraints vs. QCD-based ones, and tree-dominated observables
vs. loop-dominated ones. In
the bottom left plot the constraint on $\alpha$ has been used assuming
there is no New Physics contribution to the $\Delta I=3/2$ $b\to d$
electroweak penguin amplitudes. In the combination of this constraint with
$\beta$ from $B\to M_{c\bar c} K_S$ modes the New Physics mixing phase
cancels, so that it gives a New Physics-free determination of
$\gamma=\pi-\beta-\alpha$.}
\end{figure*}

The seven constraints in the $(\rhobar,\etabar)$ plane somewhat dilutes
the information. We have thus done partial analyses in order to compare,
respectively, CP-violating observables with CP-conserving ones,
theory-free constraints with QCD-based ones, and tree-dominated
observables with loop-dominated ones. The overall consistency of these
fits is striking, and the small deviations that can be seen here and there
are well compatible with what is expected from the
convolution of statistical
fluctuations with theoretical uncertainties.

\section{New Physics in mixing amplitudes}

\begin{figure*}[Ht]
\includegraphics[width=0.475\textwidth]{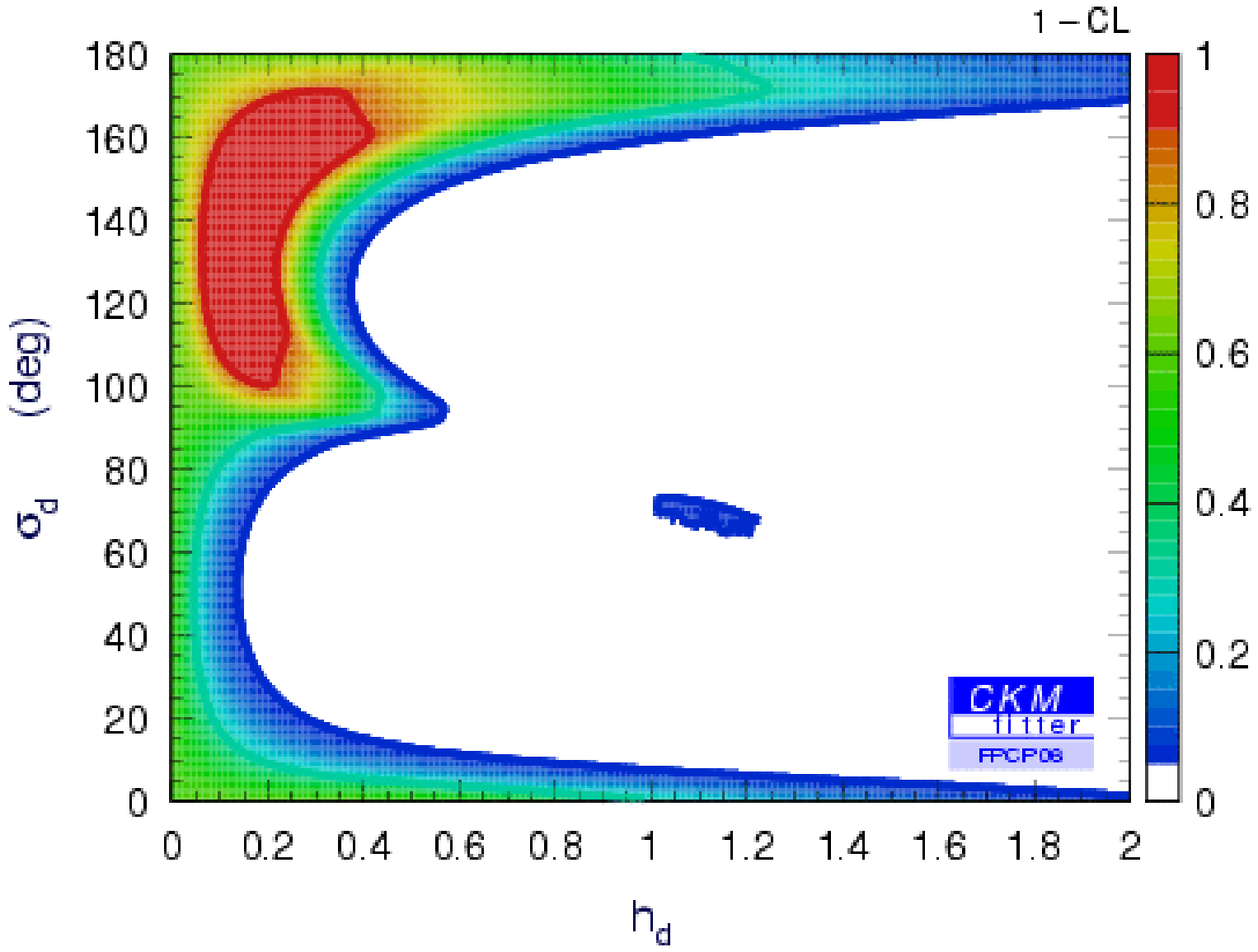}\hfill%
\includegraphics[width=0.475\textwidth]{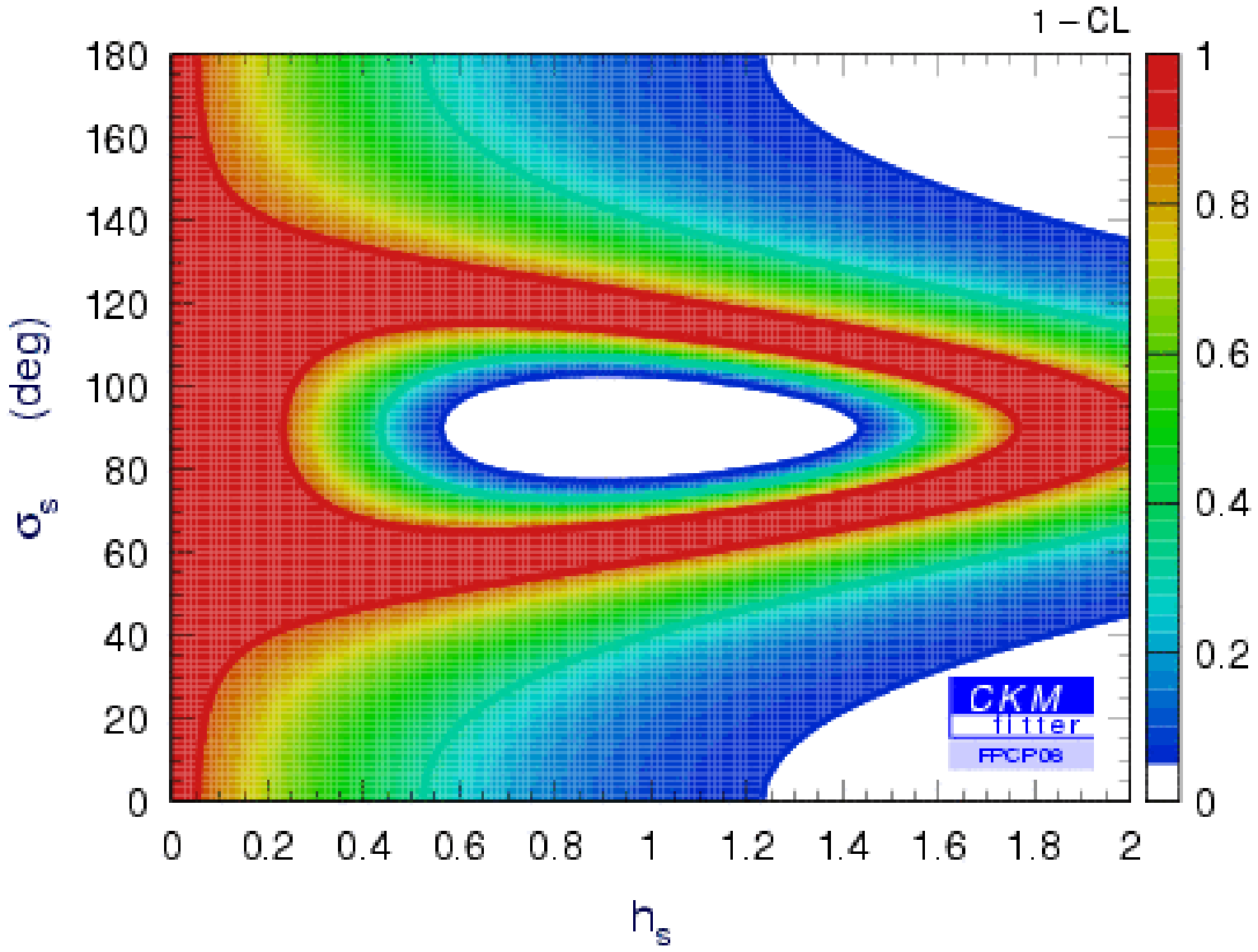}
\caption{Confidence level in the $(h,\sigma)$ plane for the global CKM fit, assuming possible New
Physics contributions to the mixing amplitudes. $(h,\sigma)$ parametrizes the deviation of the
matrix elements with respect to their Standard Model values~\cite{agashe},
so that SM
corresponds to $h=0$. In addition to the standard inputs we use the
semileptonic asymmetry $\mathcal{A}_\mathrm{SL}(B_d)=-0.0005 \pm 0.0055$.
Left: $B_d\bar{B_d}$ system; right:
$B_s\bar{B_s}$ system.}\label{NP1}
\end{figure*}

\begin{figure*}[Ht]
\centering
\includegraphics[width=0.475\textwidth]{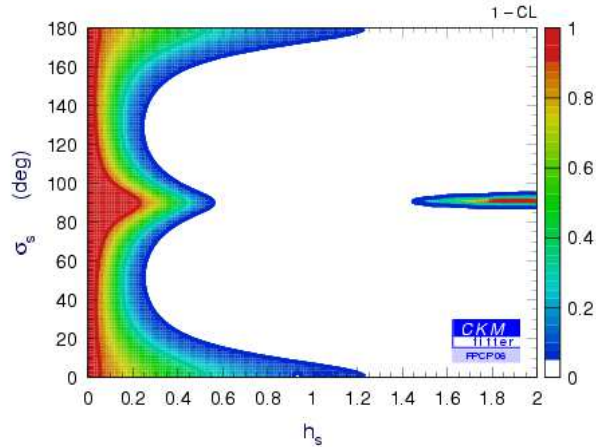}
\label{NP2}\caption{Confidence level in the $(h_s,\sigma_s)$ plane for the
global CKM fit, assuming possible New Physics contributions to the mixing
amplitudes, and a measurement of $B_s\bar{B_s}$ mixing at $\dms=20.000\pm
0.035\ \mathrm{ps}^{-1}$, $\sin 2\beta_s=0.036\pm 0.100$ that
approximately corresponds a few weeks of LHCb running ($0.2$ fb$^{-1}$).}
\end{figure*}

We have updated the model-independent analysis of possible New Physics effects in meson mixing
amplitudes, as described in Ref.~\cite{bible}
%(we use the $(h,\sigma)$ parametrization of Ref.~\cite{agashe}).
For the
$B_d\bar{B_d}$ case (Fig.~\ref{NP1}), the data
are well compatible with Standard Model values for the parameters, and the region involving new contributions
has decreased. For $B_s\bar{B_s}$ the constraint is weak, despite the new $\dms$ input: this is
because of the theoretical uncertainties. On Fig~\ref{NP2} a tentative extrapolation of the
situation after a few weeks of nominal LHCb running is shown, when the
$B_s\bar{B_s}$ mixing phase will have been
probably measured.

\section{Conclusion}
Since the last two years, CKM physics has entered its precision era. Experimental data become
more precise and systematics and specific statistical effects require
careful treatment. On the other hand 
theory of hadronic matrix elements has made progress and is expected to continue on the way. For the
``cleanest'' observables the
Standard Model does not show any sizable anomaly with respect to the data. Still, important
observables, such as very rare kaon or $B$ decays, are missing from the overall pattern and future
experiments and theoretical methods will bring up new information on quark mixing and CP violation.

\bigskip % extra skip inserted
\begin{acknowledgments}
Centre de Physique Th\'eorique is UMR 6207 du CNRS associ\'ee aux Universit\'es d'Aix-Marseille I et
II et Universit\'e du Sud Toulon-Var; laboratoire affili\'e \`a
la FRUMAM-FR2291. 

Work partially
supported by EC-Contract HPRN-CT-2002-00311 (EURIDICE).
\end{acknowledgments}

\begin{table*}[p]
\begin{center}
\caption{\label{tab:ckmInputs} \em
        Inputs to the standard CKM fit. 
        If not stated otherwise: for two errors given, the 
        first is statistical and accountable systematic and the 
        second stands for systematic theoretical uncertainties.
        The last two columns indicate\rfit\
        treatment of the input parameters:
        measurements or parameters that have statistical errors 
        (we include here experimental systematics)
        are marked in the ``GS'' column by an asterisk; measurements
        or parameters that have systematic theoretical
        errors are marked in the ``TH'' column by an asterisk.
        \underline{Upper part:} 
        experimental determinations of the CKM matrix elements.
        \underline{Middle upper part:} 
        \CP-violation and mixing observables.
        \underline{Middle lower part:} 
        parameters used in SM predictions that are 
        obtained from experiment.
        \underline{Lower part:} 
        parameters of the SM predictions obtained from 
        theory.}
\setlength{\tabcolsep}{0.0pc}
{\small
\begin{tabular*}{\textwidth}{@{\extracolsep{\fill}}lcccc}\hline
&&&& \\[-0.3cm]
        &       &       &       \mc{2}{c}{Errors}       \\
  \rs{Parameter}        & \rs{Value $\pm$ Error(s)}
                & \rs{Reference}
                & GS
                & TH \\[0.15cm]
\hline
&&&& \\
% lambda determined now by super-allowed beta-decays
% Vud from neutrons excluded for now due to neutron lifetime problem
% Vus numbers not used as these are not settled
  \rs{$\Vud$ (nuclei)}  
                & \rs{$0.97377\pm0.00027$}
                & \rs{\cite{ckm05}}
                & \rs{$\star$} & \rs{-} \\
  \rs{$\Vus$ ($K_{\ell3}$ and $K_{\mu2}$)}  
                & \rs{$0.2257\pm0.0021$}
                & \rs{\cite{PDG}}
                & \rs{$\star$} & \rs{-} \\
  \rs{$\Vub$ (incl.)}         
                & \rs{$(4.45 \pm 0.23 \pm 0.39) \times 10^{-3}$}
                & \rs{\cite{HFAG,vubcomment}}
                & \rs{$\star$} & \rs{$\star$} \\
  \rs{$\Vub$ (excl.)}         
                & \rs{$(3.94 \pm 0.28 \pm 0.51) \times 10^{-3}$}
                & \rs{\cite{PDG}}
                & \rs{$\star$} & \rs{$\star$} \\
  \rs{$\Vcb$ (incl.)}   
                & \rs{$(41.70\pm 0.70 ) \times 10^{-3}$}
                & \rs{\cite{PDG}}
                & \rs{$\star$} & \rs{-} \\
  \rs{$\Vcb$ (excl.)}   
                & \rs{$(41.18\pm 1.71)\times10^{-3}$}
                & \rs{\cite{HFAG}}
                & \rs{$\star$} & \rs{-} \\
\hline
&&&& \\
  \rs{$|\epsk|$}        
                & \rs{$(2.221\pm0.008)\times10^{-3}$}
                & \rs{\cite{KTeVKLOE}}
                & \rs{$\star$} & \rs{-} \\
  \rs{$\dmd$}           
                & \rs{$(0.507 \pm 0.004)~{\rm ps}^{-1}$}
                & \rs{\cite{HFAG}}
                & \rs{$\star$} & \rs{-} \\
  \rs{$\dms$}           
                & \rs{\footnotesize Amplitude spectrum+CDF -LogL}
                & \rs{\cite{fpcp06}}
                & \rs{$\star$} & \rs{-} \\
  \rs{$\stbwa$} & \rs{$0.687 \pm 0.032$}
                & \rs{\cite{HFAG}}
                & \rs{$\star$} & \rs{-} \\
\hline
&&&& \\
  \rs{$S_{\pi\pi}^{+-}$} & \rs{$-0.50 \pm 0.12$}
                & \rs{\cite{HFAG}}
                & \rs{$\star$} & \rs{-} \\
  \rs{$C_{\pi\pi}^{+-}$} & \rs{$-0.37 \pm 0.10$}
                & \rs{\cite{HFAG}}
                & \rs{$\star$} & \rs{-} \\
  \rs{$C_{\pi\pi}^{00}$}  & \rs{$-0.28 \pm 0.39$}
                & \rs{\cite{HFAG}}
                & \rs{$\star$} & \rs{-} \\
  \rs{$\BR_{\pi\pi}$ all charges}  & \rs{Inputs to isospin analysis}
                & \rs{\cite{HFAG}}
                & \rs{$\star$} & \rs{-} \\
\hline
&&&& \\
  \rs{$S_{\rho\rho,L}^{+-}$} & \rs{$-0.22\pm0.22$}
                & \rs{\cite{HFAG}}
                & \rs{$\star$} & \rs{-} \\
  \rs{$C_{\rho\rho,L}^{+-}$} & \rs{$-0.02 \pm 0.17$}
                & \rs{\cite{HFAG}}
                & \rs{$\star$} & \rs{-} \\
  \rs{$\BR_{\rho\rho,L}$ all charges}  & \rs{Inputs to isospin analysis}
                & \rs{\cite{HFAG}}
                & \rs{$\star$} & \rs{-} \\
\hline
&&&& \\
  \rs{$B^0 \rightarrow (\rho \pi)^0 \rightarrow 3\pi$} &
\rs{Time-dependent Dalitz analysis}
                & \rs{\cite{BABAR-dalitz3pi}}
                & \rs{$\star$} & \rs{-} \\
\hline
&&&& \\
  \rs{$B^{-}\rightarrow D^{(*)} K^{(*)-}$} & \rs{Inputs to GLW analysis}
                & \rs{\cite{HFAG}}
                & \rs{$\star$} & \rs{-} \\
  \rs{$B^{-}\rightarrow D^{(*)} K^{(*)-}$} & \rs{Inputs to ADS analysis}
                & \rs{\cite{HFAG}}
                & \rs{$\star$} & \rs{-} \\
  \rs{$B^{-}\rightarrow D^{(*)} K^{(*)-}$} & \rs{GGSZ Dalitz analyses}
                & \rs{\cite{HFAG}}
                & \rs{$\star$} & \rs{-} \\
\hline
&&&& \\
  \rs{$\BR(B^-\to\tau^-\nutb)$} & \rs{Experimental likelihoods}
            & \rs{\cite{btaunu}}
            & \rs{$\star$} & \rs{-} \\
\hline
&&&& \\
  \rs{$\mcRun(m_c)$}    & \rs{$(1.24\pm0.037\pm0.095)\gevcc$}
                & \rs{\cite{BF}}
                & \rs{$\star$} 
                & \rs{$\star$} \\
  \rs{$\mtRun(m_t)$}    
                & \rs{$(162.3\pm2.2)\gevcc$}
                & \rs{\cite{TopCDFD0Tev}}
                & \rs{$\star$} 
                & \rs{-} \\
  \rs{$m_{\Kp}$}            
                & \rs{$(493.677\pm0.016)\mevcc$}
                & \rs{\cite{PDG}}
                & \rs{-}
                & \rs{-} \\
  \rs{$\Delta m_K$}     
                & \rs{$(3.4833 \pm 0.0066)\times 10^{-12}\mevcc$}
                & \rs{\cite{PDG}}
                & \rs{-}
                & \rs{-} \\
  \rs{$m_{B_d}$}        
                & \rs{$(5.2794\pm0.0005)\gevcc$}
                & \rs{\cite{PDG}}
                & \rs{-}
                & \rs{-} \\
  \rs{$m_{B_s}$}        
                & \rs{$(5.3696\pm0.0024)\gevcc$}
                & \rs{\cite{PDG}}
                & \rs{-}
                & \rs{-} \\
  \rs{$m_W$}            
                & \rs{$(80.423\pm0.039)\gevcc$}
                & \rs{\cite{PDG}}
                & \rs{-}
                & \rs{-} \\
  \rs{$G_F$}            
                & \rs{$1.16639\times 10^{-5}\gev{}^{-2}$}
                & \rs{\cite{PDG}}
                & \rs{-}
                & \rs{-} \\
  \rs{$f_K$}            
                & \rs{$(159.8\pm1.5)\mev$}
                & \rs{\cite{PDG}}
                & \rs{-}
                & \rs{-} \\
\hline
\end{tabular*}
}
\end{center}
\end{table*}

\begin{table*}[p]
\begin{center}
\caption{Inputs to the standard CKM fit (continued)}\label{input2}
\setlength{\tabcolsep}{0.0pc}
{\small
\begin{tabular*}{\textwidth}{@{\extracolsep{\fill}}lcccc}\hline
&&&& \\[-0.3cm]
        &       &       &       \mc{2}{c}{Errors}       \\
  \rs{Parameter}        & \rs{Value $\pm$ Error(s)}
                & \rs{Reference}
                & GS
                & TH \\[0.15cm]
\hline
&&&& \\
  \rs{$B_K$}            
                & \rs{$0.79\pm0.04\pm0.09$}
                & \rs{\cite{ckm05}}
                & \rs{$\star$} 
                & \rs{$\star$} \\
  \rs{$\as(m_Z^2)$}
                & \rs{$0.1176\pm0.0020$}
                & \rs{\cite{PDG}}
                & \rs{-} 
                & \rs{$\star$} \\
  \rs{$\eta_{ct}$}      
                & \rs{$0.47\pm0.04$}
                & \rs{\cite{Nierste}}
                & \rs{-}
                & \rs{$\star$} \\
  \rs{$\eta_{tt}$}
                & \rs{$0.5765\pm0.0065$}
                & \rs{\cite{Nierste,Nierste2}}
                & \rs{-}
                & \rs{$\star$} \\
  \rs{$\etaB(\MSbar)$}
                & \rs{$0.551\pm0.007$ }
                & \rs{\cite{bbl}}
                & \rs{-}
                & \rs{$\star$} \\
%  \rs{$\fbd\sqrt{B_d}$}
%                & \rs{$(223\pm33\pm12)\mev$}
%                & \rs{\cite{ckm05}}
%                & \rs{$\star$} 
%                & \rs{$\star$} \\
  \rs{$\fbd$}
                & \rs{$(191\pm27)\mev$}
                & \rs{\cite{ckm05}}
                & \rs{$\star$} 
                & \rs{-} \\
  \rs{$B_d$}
                & \rs{$1.37\pm0.14$}
                & \rs{\cite{ckm05}}
                & \rs{$\star$} 
                & \rs{-} \\
  \rs{$\xi^{(a)}$}            
                & \rs{$1.24 \pm 0.04 \pm 0.06$}
                & \rs{\cite{ckm05}}
                & \rs{$\star$} 
                & \rs{$\star$} \\
\hline
\end{tabular*}
}
\end{center}
\vspace{-0.3cm}
{\footnotesize $^{(a)}$anticorrelated theory error with $\fbd\sqrt{B_d}$.}
\end{table*}

\begin{table*}[p]
\begin{center}
\caption{\label{output}Fit results and errors using the standard input
observables. For results
marked with ``meas. not in fit'', the measurement of the corresponding
observable has not been
included in the fit.}
 \setlength{\tabcolsep}{0.8pc}
 {\normalsize
 \begin{tabular}{lccc} \hline &&& \\[-0.3cm]
 Observable & central $\pm$ ${\rm CL}\equiv 1\sigma$ & 
 $\pm$ ${\rm CL}\equiv 2\sigma$ & $\pm$ ${\rm CL}\equiv 3\sigma$ 
\\[0.15cm]
 \hline  &&& \\[-0.3cm]
 $\lambda$                                                              &
$   0.2272 ^{\,+   0.0010}_{\,-   0.0010}$ & $^{\,+   0.0020}_{\,-  
0.0020}$ & $^{\,+   0.0030}_{\,-   0.0030}$
 \\[0.15cm]
 $A$                                                                    &
$    0.809 ^{\,+    0.014}_{\,-    0.014}$ & $^{\,+    0.029}_{\,-   
0.028}$ & $^{\,+    0.044}_{\,-    0.042}$
 \\[0.15cm]
 $\bar\rho$                                                             &
$    0.197 ^{\,+    0.026}_{\,-    0.030}$ & $^{\,+    0.050}_{\,-   
0.087}$ & $^{\,+    0.074}_{\,-    0.133}$
 \\[0.15cm]
 $\bar\eta$                                                             &
$    0.339 ^{\,+    0.019}_{\,-    0.018}$ & $^{\,+    0.047}_{\,-   
0.037}$ & $^{\,+    0.075}_{\,-    0.057}$
 \\[0.15cm]
 \hline &&&      \\[-0.3cm]
 $J$~~$[10^{-5}]$                                                       &
$     3.05 ^{\,+     0.18}_{\,-     0.18}$ & $^{\,+     0.45}_{\,-    
0.36}$ & $^{\,+     0.69}_{\,-     0.54}$
 \\[0.15cm]
 \hline &&&      \\[-0.3cm]
 $\sin(2\alpha)$                                                        &
$    -0.25 ^{\,+     0.17}_{\,-     0.15}$ & $^{\,+     0.49}_{\,-    
0.28}$ & $^{\,+     0.71}_{\,-     0.42}$
 \\[0.15cm]
 $\sin(2\alpha)$ {\small (meas. not in fit)}                            &
$    -0.23 ^{\,+     0.55}_{\,-     0.16}$ & $^{\,+     0.72}_{\,-    
0.32}$ & $^{\,+     0.83}_{\,-     0.45}$
 \\[0.15cm]
 $\sin(2\beta)$                                                         &
$    0.716 ^{\,+    0.024}_{\,-    0.024}$ & $^{\,+    0.048}_{\,-   
0.049}$ & $^{\,+    0.074}_{\,-    0.075}$
 \\[0.15cm]
 $\sin(2\beta)$ {\small (meas. not in fit)}                             &
$    0.752 ^{\,+    0.057}_{\,-    0.035}$ & $^{\,+    0.105}_{\,-   
0.073}$ & $^{\,+    0.135}_{\,-    0.112}$
 \\[0.15cm]
 $\alpha$~~(deg)                                                        &
$     97.3 ^{\,+      4.5}_{\,-      5.0}$ & $^{\,+      8.7}_{\,-    
14.0}$ & $^{\,+     13.7}_{\,-     20.7}$
 \\[0.15cm]
 $\alpha$~~(deg) {\small (meas. not in fit)}                            &
$     96.5 ^{\,+      4.9}_{\,-     16.0}$ & $^{\,+      9.9}_{\,-    
21.2}$ & $^{\,+     14.6}_{\,-     25.3}$
 \\[0.15cm]
 $\alpha$~~(deg) {\small (dir. meas.)}                                  &
$    100.2 ^{\,+     15.0}_{\,-      8.8}$ & $^{\,+     22.7}_{\,-    
18.2}$  & $^{\,+     32.0}_{\,-     28.1}$
 \\[0.15cm]
 $\beta$~~(deg)                                                         &
$    22.86 ^{\,+     1.00}_{\,-     1.00}$ & $^{\,+     2.03}_{\,-    
1.97}$ & $^{\,+     3.22}_{\,-     2.93}$
 \\[0.15cm]
 $\beta$~~(deg) {\small (meas. not in fit)}                             &
$     24.4 ^{\,+      2.6}_{\,-      1.5}$ & $^{\,+      5.1}_{\,-     
3.0}$ & $^{\,+      6.9}_{\,-      4.5}$
 \\[0.15cm]
 $\beta$~~(deg) {\small (dir. meas.)}                                   &
$     21.7 ^{\,+      1.3}_{\,-      1.2}$ & $^{\,+      2.6}_{\,-     
2.4}$& $^{\,+      4.1}_{\,-      3.6}$
 \\[0.15cm]
 $\gamma\simeq\delta$~~(deg)                                            &
$     59.8 ^{\,+      4.9}_{\,-      4.1}$ & $^{\,+     13.9}_{\,-     
7.9}$ & $^{\,+     20.8}_{\,-     12.1}$
 \\[0.15cm]
 $\gamma\simeq\delta$~~(deg) {\small (meas. not in fit)}                &
$     59.8 ^{\,+      4.9}_{\,-      4.2}$ & $^{\,+     14.1}_{\,-     
8.0}$ & $^{\,+     21.0}_{\,-     12.3}$
 \\[0.15cm]
 $\gamma\simeq\delta$~~(deg) {\small (dir. meas.)}                      &
$       63 ^{\,+       35}_{\,-       25}$ & $^{\,+       62}_{\,-      
40}$& $^{\,+     100}_{\,-     54}$
 \\[0.15cm]
 \hline &&&      \\[-0.3cm]
 $\beta_s$~~(deg)                                                       &
$    1.045 ^{\,+    0.061}_{\,-    0.057}$ & $^{\,+    0.151}_{\,-   
0.114}$ & $^{\,+    0.238}_{\,-    0.177}$
 \\[0.15cm]
 $\sin(2\beta_s)$                                                       &
$   0.0365 ^{\,+   0.0021}_{\,-   0.0020}$ & $^{\,+   0.0053}_{\,-  
0.0040}$ & $^{\,+   0.0083}_{\,-   0.0062}$
 \\[0.15cm]
 \hline &&&      \\[-0.3cm]
 $R_u$                                                                  &
$    0.391 ^{\,+    0.015}_{\,-    0.015}$ & $^{\,+    0.031}_{\,-   
0.029}$ & $^{\,+    0.049}_{\,-    0.044}$
 \\[0.15cm]
 $R_t$                                                                  &
$    0.872 ^{\,+    0.033}_{\,-    0.028}$ & $^{\,+    0.095}_{\,-   
0.054}$ & $^{\,+    0.143}_{\,-    0.082}$
 \\[0.15cm]
 \hline &&&      \\[-0.3cm]
 $\Delta m_d$~~$({\rm ps}^{-1})$ {\small (meas. not in fit)}            &
$    0.394 ^{\,+    0.097}_{\,-    0.097}$ & $^{\,+    0.219}_{\,-   
0.132}$ & $^{\,+    0.361}_{\,-    0.162}$
 \\[0.15cm]
 $\Delta m_s$~~$({\rm ps}^{-1})$                                        &
$    17.34 ^{\,+     0.49}_{\,-     0.20}$ & $^{\,+     0.65}_{\,-    
0.35}$ & $^{\,+     0.78}_{\,-     0.49}$
 \\[0.15cm]
 $\Delta m_s$~~$({\rm ps}^{-1})$  {\small (meas. not in fit)}           &
$     21.7 ^{\,+      5.9}_{\,-      4.2}$ & $^{\,+      9.7}_{\,-     
6.8}$ & $^{\,+     13.1}_{\,-      9.1}$
 \\[0.15cm]
 $\epsilon_K$~~$[10^{-3}]$ {\small (meas. not in fit)}                  &
$     2.46 ^{\,+     0.63}_{\,-     0.88}$ & $^{\,+     1.05}_{\,-    
1.05}$ & $^{\,+     1.50}_{\,-     1.20}$
 \\[0.15cm]
 \hline &&&      \\[-0.3cm]
 $f_{B_d}$~~(MeV) {\small (lattice value not in fit)}                   &
$      183 ^{\,+       10}_{\,-       10}$ & $^{\,+       21}_{\,-      
20}$ & $^{\,+       34}_{\,-       28}$
 \\[0.15cm]
 $\xi$ {\small (lattice value not in fit)}     & $    1.061 ^{\,+   
0.122}_{\,-    0.047}$ & $^{\,+    0.213}_{\,-    0.083}$ & $^{\,+   
0.324}_{\,-    0.119}$
 \\[0.15cm]
 $B_K$ {\small (lattice value not in fit)}                              &
$    0.722 ^{\,+    0.251}_{\,-    0.084}$ & $^{\,+    0.348}_{\,-   
0.157}$ & $^{\,+    0.461}_{\,-    0.216}$
 \\[0.15cm]
 \hline &&&      \\[-0.3cm]
 $m_c$~~$({\rm GeV}/c^2)$ {\small (meas. not in fit)}                   &
$     0.81 ^{\,+     0.93}_{\,-     0.36}$ & $^{\,+     1.08}_{\,-    
0.36}$ & $^{\,+     1.23}_{\,-     0.81}$
 \\[0.15cm]
 $m_t$~~$({\rm GeV}/c^2)$ {\small (meas. not in fit)}                   &
$      150 ^{\,+       27}_{\,-       21}$ & $^{\,+       57}_{\,-      
35}$ & $^{\,+       79}_{\,-       46}$
 \\[0.15cm]
 \hline
 \end{tabular}
 }
 \end{center}
\end{table*}

\begin{table*}[p]
\caption{Fit results (continued)}\label{output2}
\begin{center}
 \setlength{\tabcolsep}{0.8pc}
 {\normalsize
 \begin{tabular}{lccc} \hline &&& \\[-0.3cm]
 Quantity & central $\pm$ ${\rm CL}\equiv 1\sigma$ & 
 $\pm$ ${\rm CL}\equiv 2\sigma$ & $\pm$ ${\rm CL}\equiv 3\sigma$ 
\\[0.15cm]
 \hline  &&& \\[-0.3cm]
 $\Bptaun$~~$[10^{-5}]$                                                 &
$      9.6 ^{\,+      1.5}_{\,-      1.5}$ & $^{\,+      3.3}_{\,-     
2.9}$ & $^{\,+      5.4}_{\,-      4.0}$
 \\[0.15cm]
 $\Bpmun$~~$[10^{-7}]$                                                  &
$     4.32 ^{\,+     0.58}_{\,-     0.57}$ & $^{\,+     1.27}_{\,-    
1.12}$ & $^{\,+     2.05}_{\,-     1.62}$
 \\[0.15cm]
 \hline &&&      \\[-0.3cm]
 $\Kzpiznn$~~$[10^{-11}]$                                               &
$     2.58 ^{\,+     0.48}_{\,-     0.40}$ & $^{\,+     1.01}_{\,-    
0.68}$ & $^{\,+     1.53}_{\,-     0.93}$
 \\[0.15cm]
 $\Kppipnn$~~$[10^{-11}]$                                               &
$      7.5 ^{\,+      1.8}_{\,-      2.0}$ & $^{\,+      2.5}_{\,-     
2.4}$ & $^{\,+      3.2}_{\,-      2.7}$
 \\[0.15cm]
 \hline &&&      \\[-0.3cm]
 $|V_{ud}|$                                                             &
$  0.97383 ^{\,+  0.00024}_{\,-  0.00023}$ & $^{\,+  0.00047}_{\,- 
0.00047}$ & $^{\,+  0.00071}_{\,-  0.00071}$
 \\[0.15cm]
 $|V_{us}|$                                                             &
$   0.2272 ^{\,+   0.0010}_{\,-   0.0010}$ & $^{\,+   0.0020}_{\,-  
0.0020}$ & $^{\,+   0.0030}_{\,-   0.0030}$
 \\[0.15cm]
 $|V_{ub}|$~~$[10^{-3}]$                                                &
$     3.82 ^{\,+     0.15}_{\,-     0.15}$ & $^{\,+     0.31}_{\,-    
0.29}$ & $^{\,+     0.49}_{\,-     0.44}$
 \\[0.15cm]
 $|V_{ub}|$~~$[10^{-3}]$ {\small (meas. not in fit)}                    &
$     3.64 ^{\,+     0.19}_{\,-     0.18}$ & $^{\,+     0.39}_{\,-    
0.36}$ & $^{\,+     0.60}_{\,-     0.55}$
 \\[0.15cm]
 $|V_{cd}|$                                                             &
$  0.22712 ^{\,+  0.00099}_{\,-  0.00103}$ & $^{\,+  0.00199}_{\,- 
0.00205}$ & $^{\,+  0.00300}_{\,-  0.00307}$
 \\[0.15cm]
 $|V_{cs}|$                                                             &
$  0.97297 ^{\,+  0.00024}_{\,-  0.00023}$ & $^{\,+  0.00048}_{\,- 
0.00047}$ & $^{\,+  0.00071}_{\,-  0.00071}$
 \\[0.15cm]
 $|V_{cb}|$~~$[10^{-3}]$                                                &
$    41.79 ^{\,+     0.63}_{\,-     0.63}$ & $^{\,+     1.26}_{\,-    
1.27}$ & $^{\,+     1.89}_{\,-     1.90}$
 \\[0.15cm]
 $|V_{cb}|$~~$[10^{-3}]$ {\small (meas. not in fit)}                    &
$     44.9 ^{\,+      1.2}_{\,-      2.8}$ & $^{\,+      2.4}_{\,-     
5.7}$ & $^{\,+      3.8}_{\,-      7.7}$
 \\[0.15cm]
 $|V_{td}|$~~$[10^{-3}]$                                                &
$     8.28 ^{\,+     0.33}_{\,-     0.29}$ & $^{\,+     0.92}_{\,-    
0.57}$ & $^{\,+     1.38}_{\,-     0.86}$
 \\[0.15cm]
 $|V_{ts}|$~~$[10^{-3}]$                                                &
$    41.13 ^{\,+     0.63}_{\,-     0.62}$ & $^{\,+     1.25}_{\,-    
1.24}$ & $^{\,+     1.87}_{\,-     1.86}$
 \\[0.15cm]
 $|V_{tb}|$                                                             &
$ 0.999119 ^{\,+ 0.000026}_{\,- 0.000027}$ & $^{\,+ 0.000052}_{\,-
0.000054}$ & $^{\,+ 0.000078}_{\,- 0.000082}$
 \\[0.15cm]
 \hline &&&      \\[-0.3cm]
 $|V_{td}/V_{ts}|$                                                      &
$   0.2011 ^{\,+   0.0081}_{\,-   0.0065}$ & $^{\,+   0.0230}_{\,-  
0.0127}$ & $^{\,+   0.0345}_{\,-   0.0195}$
 \\[0.15cm]
 $|V_{ud}V_{ub}^*|$~~$[10^{-3}]$                                        &
$     3.72 ^{\,+     0.15}_{\,-     0.14}$ & $^{\,+     0.30}_{\,-    
0.29}$ & $^{\,+     0.48}_{\,-     0.43}$
 \\[0.15cm]
 $\arg\left[V_{ud}V_{ub}^*\right]$~~(deg)                               &
$     59.8 ^{\,+      4.9}_{\,-      4.0}$ & $^{\,+     13.9}_{\,-     
7.8}$ & $^{\,+     20.9}_{\,-     12.1}$
 \\[0.15cm]
 $\arg\left[-V_{ts}V_{tb}^*\right]$~~(deg)                              &
$    1.043 ^{\,+    0.061}_{\,-    0.057}$ & $^{\,+    0.151}_{\,-   
0.114}$ & $^{\,+    0.238}_{\,-    0.176}$
 \\[0.15cm]
 $|V_{cd}V_{cb}^*|$~~$[10^{-3}]$                                        &
$     9.49 ^{\,+     0.15}_{\,-     0.15}$ & $^{\,+     0.30}_{\,-    
0.30}$ & $^{\,+     0.45}_{\,-     0.45}$
 \\[0.15cm]
 $\arg\left[-V_{cd}V_{cb}^*\right]$~~(deg)                              &
$   0.0339 ^{\,+   0.0021}_{\,-   0.0020}$ & $^{\,+   0.0050}_{\,-  
0.0040}$ & $^{\,+   0.0077}_{\,-   0.0060}$
 \\[0.15cm]
 $|V_{td}V_{tb}^*|$~~$[10^{-3}]$                                        &
$     8.27 ^{\,+     0.33}_{\,-     0.29}$ & $^{\,+     0.93}_{\,-    
0.57}$ & $^{\,+     1.38}_{\,-     0.85}$
 \\[0.15cm]
 $\arg\left[V_{td}V_{tb}^*\right]$~~(deg)                               &
$   -22.84 ^{\,+     1.00}_{\,-     0.99}$ & $^{\,+     1.98}_{\,-    
2.02}$ & $^{\,+     2.93}_{\,-     3.21}$
 \\[0.15cm]
 \hline &&&      \\[-0.3cm]
 ${\rm Re}\lambda_c$                                                    &
$ -0.22098 ^{\,+  0.00095}_{\,-  0.00091}$ & $^{\,+  0.00188}_{\,- 
0.00184}$ & $^{\,+  0.00282}_{\,-  0.00275}$
 \\[0.15cm]
 ${\rm Im}\lambda_c$~~$[10^{-4}]$                                       &
$   -1.377 ^{\,+    0.080}_{\,-    0.084}$ & $^{\,+    0.161}_{\,-   
0.203}$ & $^{\,+    0.244}_{\,-    0.310}$
 \\[0.15cm]
 \hline &&&      \\[-0.3cm]
 ${\rm Re}\lambda_t$~~$[10^{-4}]$                                       &
$    -3.11 ^{\,+     0.13}_{\,-     0.14}$ & $^{\,+     0.26}_{\,-    
0.36}$ & $^{\,+     0.39}_{\,-     0.57}$
 \\[0.15cm]
 ${\rm Im}\lambda_t$~~$[10^{-4}]$                                       &
$    1.377 ^{\,+    0.084}_{\,-    0.080}$ & $^{\,+    0.203}_{\,-   
0.161}$ & $^{\,+    0.310}_{\,-    0.244}$
 \\[0.15cm]
 \hline
 \end{tabular}
 }
 \end{center}
 \end{table*}


\begin{thebibliography}{99}
\bibitem{vubcomment}    For the inclusive average we are taking the BLNP number.
                        The DGE result is very close to. The theoretical error
                        on the inclusive average is obtained by adding linearly
                        the contributions from weak annihilation, subleading
                        shape functions and the HQE uncertainty on $m_{b}$.
\end{thebibliography}
\end{document}